\documentclass[
aps,
pre,
twocolumn,
groupedaddress]{revtex4-2}

\usepackage{graphicx}  
\usepackage{dcolumn}   
\usepackage{bm}        
\usepackage{amssymb}   
\usepackage{siunitx}   
\usepackage{url}

\usepackage{subcaption}
\usepackage{float}

\begin{document}

\title{Anomalous Diffusion in a Monolayer of Lightweight Spheres Fluidized in Airflow}
\author{Shihori Koyama}
\email[]{Present address: Toyota Central R\&D Labs., Inc. Bunkyo-ku, Tokyo 112-0004, Japan}
\affiliation{Department of Aeronautics and Astronautics, Kyoto University, Kyoto 615-8540, Japan}
\author{Tomoki Matsuno}
\email[]{Present address: Sony Global Manufacturing \& Operations Corporation, Tokyo 108-0075, Japan}
\affiliation{Department of Aeronautics and Astronautics, Kyoto University, Kyoto 615-8540, Japan}
\author{Takashi Noguchi}
\email[]{Present address: Department of Mechanical Engineering, Doshisha University, Kyoto 610-0394, Japan \\
 Corresponding author: noguchi.takashi.7w@kyoto-u.jp}
\affiliation{Department of Aeronautics and Astronautics, Kyoto University, Kyoto 615-8540, Japan}
\date{\today}

\def\citesupplementalmaterial{\cite{[{See Supplemental Material at }][{ for the details of experimental apparatus, the background airflow, and the radial distribution functions for the particles.}]SupplementalMaterial}}

\begin{abstract}
  This paper presents statistical analyses of random motions in a single layer of fluidized lightweight spherical particles.
  Foam polystyrene spheres were driven by an upward airflow through the sieve mesh, and their two-dimensional motion was acquired using image analysis.
  In the bulk region, the particle velocity distributions changed from Gaussian to heavy-tailed distribution as the bulk packing fraction $\phi_b$ was increased.
  The mean square displacement of the particles exhibited transition to subdiffusion at much lower $\phi_b$ than observed in previous studies using similar setup but with heavier particles.
  A slight superdiffusion and significant growth of the correlation length in the two-body velocity correlation was observed at further large $\phi_b$.
  The effect of the wall on the dynamics of the particles was also investigated and the anisotropy of the granular temperature was found to be a useful index to discriminate between the wall region and the bulk.
  The turbulence statistics in the wake of a particle indicated a strong wall-normal asymmetry of aerodynamic forcing as the ``thermal'' agitation in the wall region.

\end{abstract}



\maketitle

\section{Introduction}

Granular systems can be fluidized when it is placed in an upward fluid flow so as
to each constituting particle is levitated and agitated by turbulent aerodynamic force.
Meanwhile, the particles lose their kinetic energy through rolling friction, fluid viscosity,
and inelastic mutual collisions as well as the negative part of work done by the aerodynamic forces.
Fluidized systems are thus in a subtle dynamic state in which perpetual aerodynamic
agitation is required to compensate for the energy loss in order for particles to behave collectively like a fluid.

The aerodynamic force acting on the particles includes the force due to the flow through the inter-particle gap and that due to the vortices in the wake of the particle \cite{Achenbach1974,Sakamoto1990}.
When the number of particles is large, the aerodynamic force and the collisions between particles are considered to be sufficiently complicated and it is often hoped that the agitation is gaussian.
This leads to an analogy to the thermal system despite the dissipative and macroscopic dynamics in the many-particle system.
Actually many studies in recent years have found that various macroscopic, strongly dissipative (and thus out-of-equilibrium) systems such as particles agitated by turbulent airflow \cite{Ojha2004,Ojha2005,Abate2006,Beverland2011,Abkenar2013,Born2017} or by mechanical vibration \cite{DAnna2003,Mayor2005,Combs2008}, can mimic Brownian behavior to an appreciable degree which has long been considered mainly in the context of energetically closed system.
Thus far a variety of studies on macroscopic two-dimensional systems have been conducted \cite{Abkenar2013,Briand2018,Lopezcastano2019,Lopezcastano2020}.

Active focus of research in this field is the behavior of densely packed particles.
As the packing fraction $\phi$ increases, the particles increasingly hamper mutual movements and eventually part of the system undergoes a transition from liquid-like to solid-like behaviors, which is called the jamming transition \cite{Zhang2005,Majmudar2007,Biroli2011}.
An association between the jamming transition and the glass transition due to cooling has been suggested \cite{Silbert2002,Keys2007,Mari2009,Parisi2010,Charbonneau2017}.
The process toward the transition gives rise to arrests of particles, i.e. the caging \cite{Philipse2003,Reis2007}, which can occur at $\phi$ well before reaching the random packing limit.
When caging occurs, a plateau appears in the mean square displacement (MSD) as a result of inhibited motion due to the particle confinement.
Essentially the same phenomenon is observed in microscopic systems such as colloidal particles \cite{Weeks2000}, and macroscopic bubble raft systems \cite{Debregeas2001,Lauridsen2004}.
In these systems, once occurred, the caging does not relax at larger $\phi$ \cite{Abate2006,Reis2007} although it may do so after very long time \cite{Abate2007}.
Moreover, in the densely packed state the individual particle is known to exhibit superdiffusion \cite{Bouchaud1990}, which is typically observed in colloids and glassy materials \cite{Zaccarelli2009,Jeon2013}.

Here we report our laboratory observation of the properties of granular flow in a monolayer of spherical particles which are fluidized two-dimensionally by airflow.
In the experiments, we used spheres made of foam polystyrene which has much lower material density (only half of a ping-pong ball) than those had been used in previous studies: Smaller density makes the particles more susceptible to aerodynamic forces \cite{Schouveiler2004,Pinar2013} and thus we expect the contactless interaction via aerodynamic forces to take more control over particle dynamics than the direct interaction via collisions.
We observed the behavior of the system for various $\phi$ at two different airspeeds which were slower than the terminal velocity of the particles.
We also examined the effect of solid walls on the ``thermal motion'' of the particles and found that velocity variance of the particles near the wall showed large anisotropy, which provides a useful discriminator of the wall boundary layer from the bulk.

\section{Experimental apparatus and method}

\begin{figure}[hbt!]
\begin{center}
\includegraphics[width=0.3\linewidth]{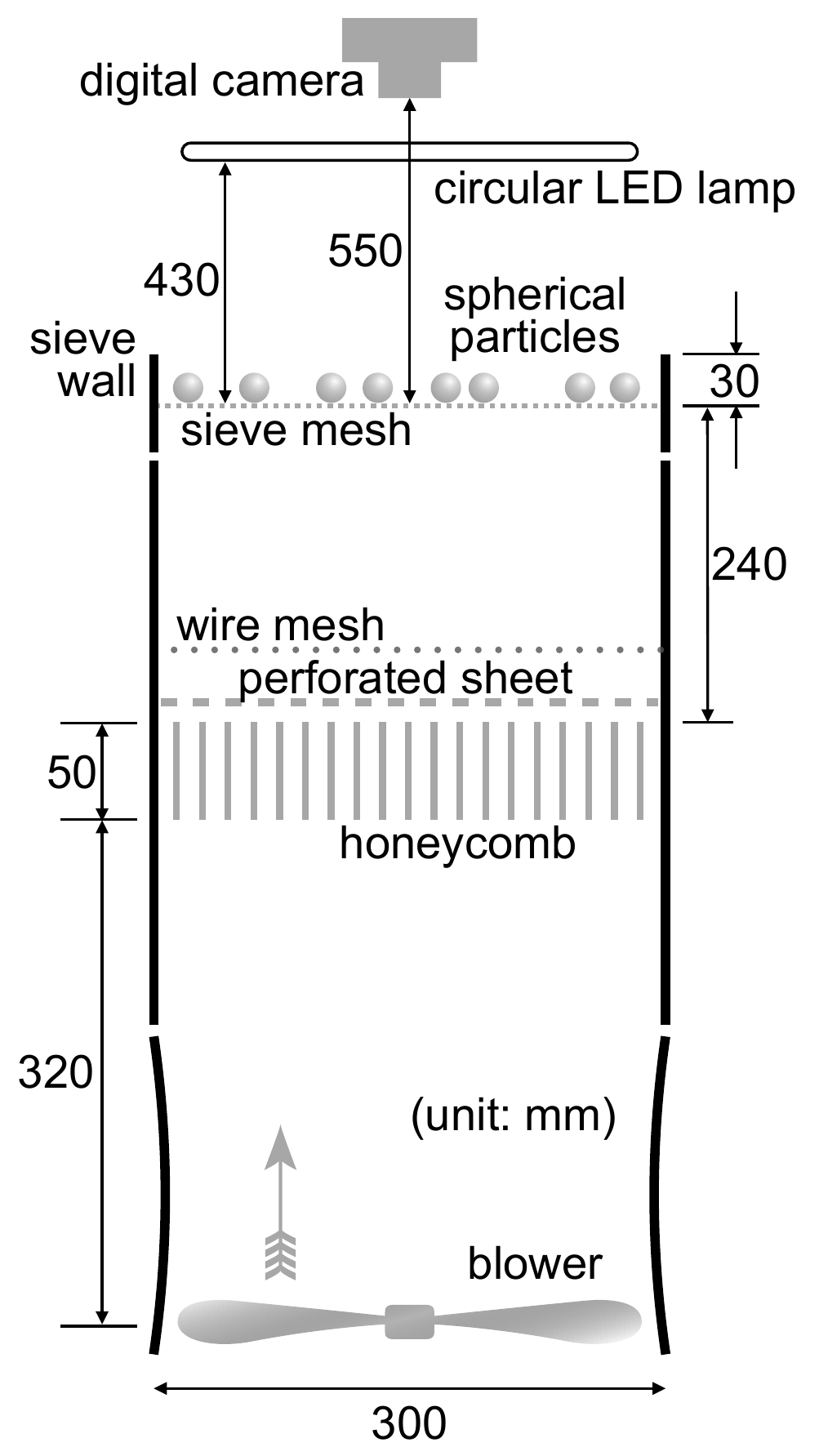} 
\end{center}
 \caption{Schematic of the experimental apparatus (not to scale). }
 \label{fig:apparatus}
\end{figure}

Spherical particles 
were placed on the mesh of a sieve and were fluidized by an upward airflow perpendicular to the plane of the mesh (Fig.~\ref{fig:apparatus}).
We used the stainless steel sieve of inner diameter of $2R_{\textrm{sieve}}= \SI{300}{mm}$,
with a side wall of \SI{30}{mm} high, and with a wire mesh spacing of \SI{150}{\micro m}.
The sieve was placed on the top of a \SI{50}{cm} long glass cylinder.
An axial fan blower (\SI{0.55}{kW} max) 
send airflow directly into the cylinder,
the halfway of which flow straighteners are installed to ensure the uniformity of airflow.

The particles we used were particles made of foamed polystyrene of radius $R_{\textrm{particle}} = \SI{10}{mm}$ and mass $M_{\textrm{particle}} = \SI{0.175}{g}$ (hence monodispersed).
The particle's terminal velocity $V_t = \SI{4.25}{m . s^{-1}}$ was obtained from its free fall in still air.
We investigated the behavior of the particles for various airspeeds $U_a$ and packing fractions $\phi$, which was controlled by the total numbers of particles $N$.
The blower power was set to two levels,
which give airflows of $U_a = (0.42 \pm 0.02) V_t$ and $(0.47 \pm 0.02) V_t$, when no particles in the sieve.
The actual airspeeds with the particle is slightly lower than these values, but could not be confidently measured by the hot-wire due to the reversal of flow associated with the turbulent wake of moving particles.

The lightness of the particles imposed tight limitations on $N$ as well as on $U_a$.
Both $N$ and $U_a$ should be small to avoid the particles to levitate higher than half the diameter especially in collision events.
At the same time, $N$ and $U_a$ should not be too small, as otherwise the resulting immobility of particles introduces a dependency on the initial configuration into the long-time behavior of the particles.
Our experiments were run within this limited range of parameters, as listed in Table \ref{tab:packingfraction}.

The effect of static charge built up on the surface of the particles was considered negligible.
Previous similar experiments \cite{Baxter2003,Baxter2007} particles underwent vibrational (thus frictional) agitation in a shallow container covered with a plexiglas lid which is coated with conducting film to prevent charging.
In our experiments the foam polystyrene particles roll always in contact with the stainless sieve mesh,
which is an unpainted metal surface.

\begin{figure}[ht]
\begin{center}
\includegraphics[width=0.3\linewidth]{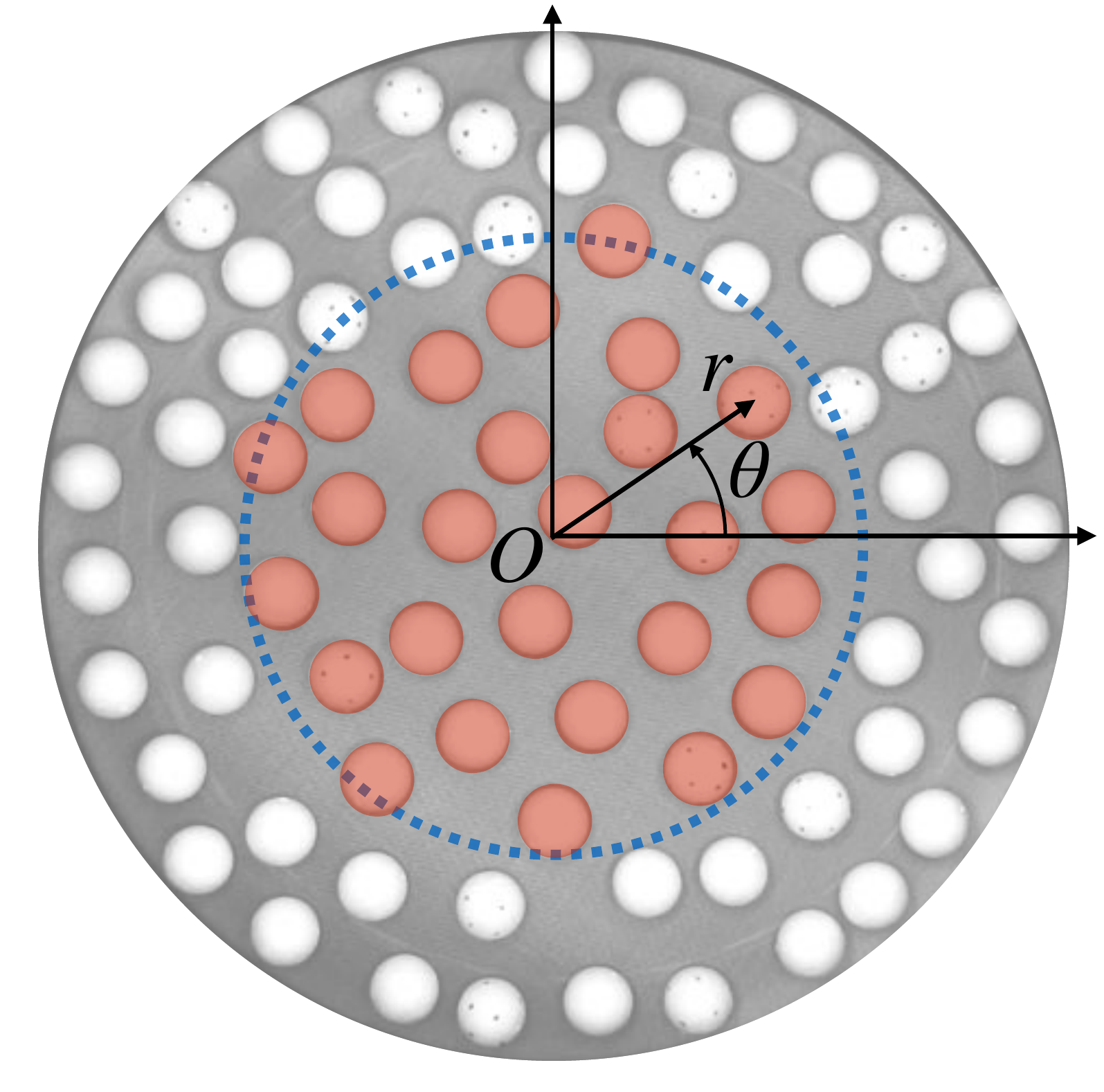} 
\end{center}
 \caption{Coordinate system used in the analysis.
 A representative image of particles ($N=80$) is displayed in background.
 Dotted circle shows the boundary between wall region and bulk region which will be defined in Section \ref{subsec:velvardist} and particles in the bulk are marked with a shade.}
 \label{fig:coordinate}
\end{figure}

We obtained the in-plane motion of all particles using an image analysis technique.
The particle motion was recorded with a digital movie camera
located on the central axis \SI{550}{mm} above (downstream) of the sieve.
The movies were recorded at 30 fps with $1920 \times 1080$ pixels for 35 min.
To minimize errors in positional detection of particle, a circular LED lamp
just above the sieve was used so that particles were uniformly illuminated.
Figure \ref{fig:coordinate} defines the laboratory coordinate system with the origin at the center of the sieve.

We analyzed the experimental movies with an in-house code using the image processing library OpenCV 3.2.0.
The code 
captures the
frames (approximately $6.3\times10^4$ frames per run),
and detects the positions of all the particles in each frame using a template matching algorithm.
The time sequence of positions for each particle was obtained by tracing the shortest-distance pair of particles between the successive frames.
The velocity and acceleration of a particle were calculated by fitting positions versus time to a third-order polynomial, with a fitting window of $\pm 4$ points and a Gaussian weighting with tapering to zero at the edges.

\section{Results}

\subsection{Density distribution}

\begin{figure}[hbt!]
  \centering
  \includegraphics[width=\linewidth]{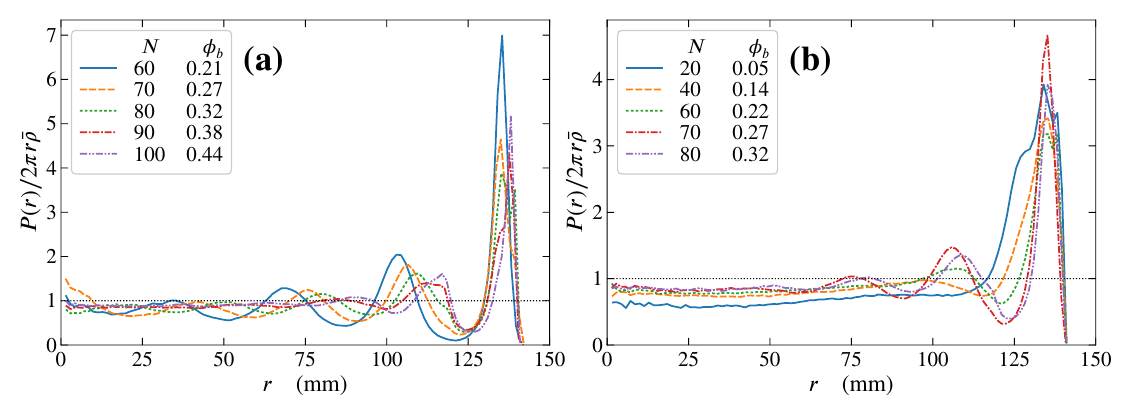}
  \caption{Spatial density distributions of particles for the various total particle numbers $N$, normalized by the mean number densities in the effective sieve radius $R_{\textrm{sieve}} - R_{\textrm{particle}} = \SI{140}{mm}$, at upflow velocities (a) $U_a = 0.42 V_t$ and (b) $U_a = 0.47 V_t$.
  See Table~\ref{tab:packingfraction} for $\phi_b$ shown in the legend. }
  \label{fig:position}
\end{figure}

The spatial density distribution using the experimental data is shown in Fig.~\ref{fig:position}.
In all cases, a significant fraction of particles tended to swarm near the wall of the sieve, which hereinafter referred to as the {\em wall region}.
In contrast, in the region near the center of the sieve, hereafter referred to as the {\em bulk region} whose precise definition will be given in the following subsection, the sieve-radial distribution approached a finite value with decaying oscillation toward the center of the sieve.

When $U_a = 0.42 V_t$ and $N = 60$, the alignment of the particles was seen from the wall to the center of the sieve surface.
When the particle number $N$ was increased, the alignment near the center became increasingly less clear and no distinct structure was visible at $N=100$.
On the other hand, when $U_a = 0.47 V_t$, the alignment was localized in the wall region when $N$ was small.
The range of the aligned particle extended toward the interior region when $N$ was increased.

\subsection{Velocity variance distribution}\label{subsec:velvardist}

\begin{figure}[hbt!]
  \centering
  \includegraphics[width=\linewidth]{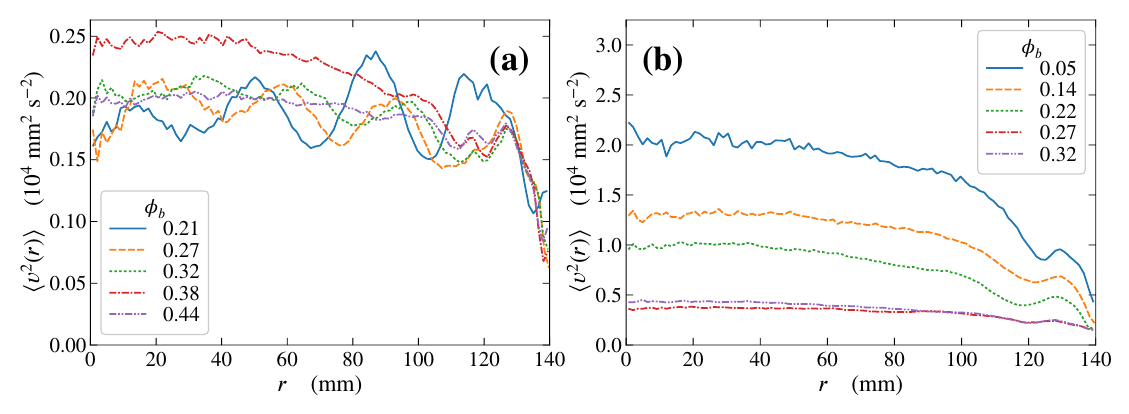}
\caption{Spatial distribution of (total) granular temperature distribution $\langle v^2(r)\rangle$, for (a) $U_a = 0.42 V_t$ and (b) $U_a = 0.47 V_t$.}
\label{fig:thermal}
\end{figure}

\begin{figure}[hbt!]
  \centering
  \includegraphics[width=\linewidth]{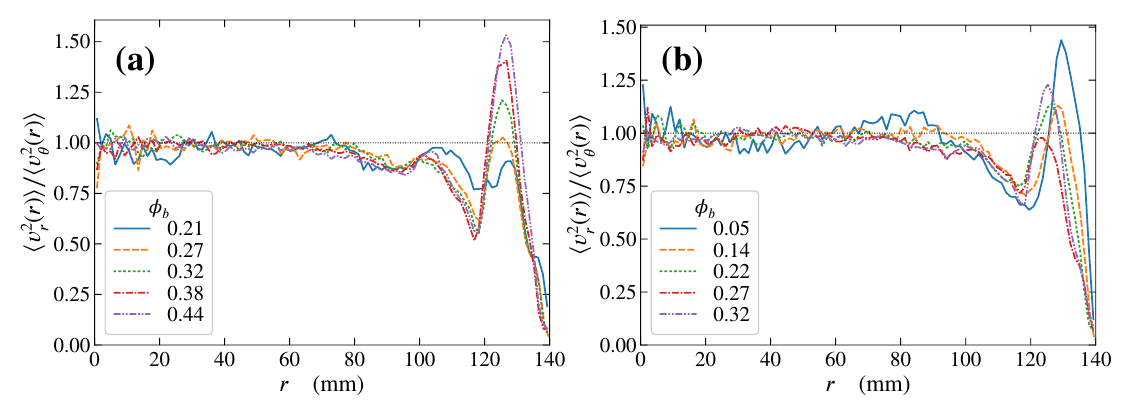}
  \caption{Spatial distribution of ratio of radial and azimuthal granular temperatures $\langle v_r^2(r)\rangle/\langle v_{\theta}^2(r)\rangle$, for (a) $U_a = 0.42 V_t$ and (b) $U_a = 0.47 V_t$.}
  \label{fig:thermalratio}
\end{figure}

Next we look at the sieve-radial distribution of velocity variance which is characteristic of the random motion of the particles, shown  in Fig.~\ref{fig:thermal}.
The velocity variance $\langle v^2\rangle$ is called the granular temperature \cite{Abate2008} due to the analogy with the thermodynamic temperature of gas molecules.
The velocity variance were uniform near the center but had a large gradient in about two particle-diameter neighborhood of the wall.
Furthermore, we obtained distributions of radial ($r$) and azimuthal ($\theta$) directional variance by decomposing the components of the particle motions.
Figure \ref{fig:thermalratio} shows the radial distribution of the ratio of radial to azimuthal variance $\langle v_r^2\rangle/\langle v_{\theta}^2\rangle$.
The random motion of the particles was anisotropic near the wall, while the motion was isotropic beyond about two particle diameter from the sieve wall.
This anisotropy contrasts the the bulk region with the wall boundary layer.
In the following discussions, we will focus on the phenomena in the bulk region by defining the boundary between this region and the wall region as the circle of \SI{90}{mm} radius from the center of the sieve.
The time averaged packing fractions in the bulk region $\phi_b$ for various $U_a$ and $N$ are listed in the Table \ref{tab:packingfraction}.
Note that since the partitioning of the number of particles between the bulk and the wall region varies with $U$, $\phi_b$ can differ even for the same $N$.

\begin{table}
\caption{Effective packing fraction in the bulk region, $\phi_b$, for various airspeeds $U_a$ (normalized by terminal velocity of particle $V_t$) and total number of particles $N$.
Since the region is open, the effective packing fraction fluctuates with time:
A fluctuation of $0.012$ is equivalent to one particle's entry or exit.
The parameter $\alpha$, which is obtained by fitting generalized exponential distribution $\exp(-|v_r|^{\alpha})$ against observed velocity distribution, is discussed in Section \ref{res:vdf}.}
\begin{ruledtabular}
\begin{tabular}{rrrr}
$U_a/V_t$  & $N$ & $\phi_b\;$ (Mean $\pm$ SD) & $\alpha $ \\
\hline
0.42  & 60 & 0.211 $\pm$ 0.015 & 1.35 \\
      & 70 & 0.269 $\pm$ 0.015 & 1.38 \\
      & 80 & 0.322 $\pm$ 0.016 & 1.56 \\
      & 90 & 0.379 $\pm$ 0.017 & 1.69 \\
     & 100 & 0.437 $\pm$ 0.018 & 1.53 \\
0.47  & 20 & 0.052 $\pm$ 0.016 & 2.29 \\
      & 40 & 0.139 $\pm$ 0.019 & 2.12 \\
      & 60 & 0.224 $\pm$ 0.019 & 2.14 \\
      & 70 & 0.274 $\pm$ 0.016 & 1.80 \\
      & 80 & 0.325 $\pm$ 0.017 & 1.65 \\
\end{tabular}
\end{ruledtabular}
\label{tab:packingfraction}
\end{table}

\subsection{MSD}

\begin{figure}[hbt!]
  \centering
  \includegraphics[width=\linewidth]{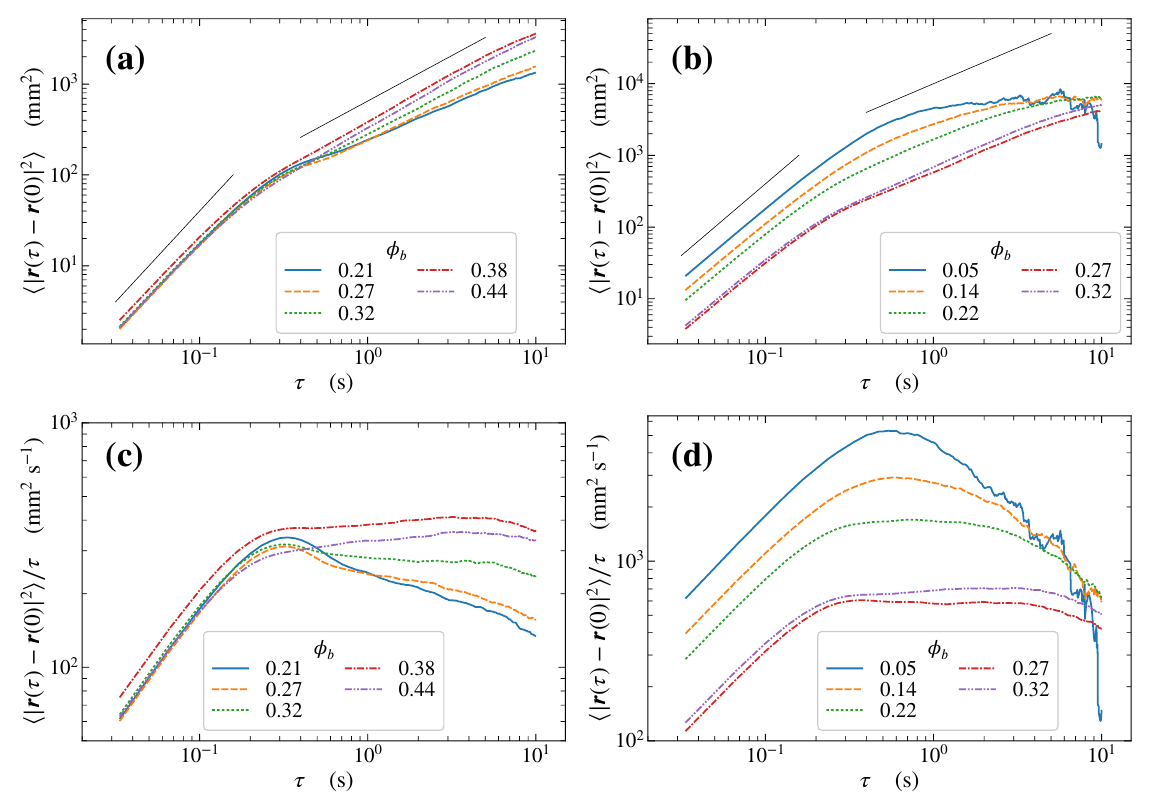}
  \caption{(a),(b) MSD of particles in the bulk region, versus time. Lines with slopes of 2 and 1 are shown as references.
  Saturation at around \SI{6e3}{mm^2} is due to the limited size ($r \le \SI{90}{mm}$) of the observation domain (the bulk region).
  (c),(d) same data as in (a) and (b) but divided by $\tau$ to give diffusion coefficients.
  Airflow velocities are (a),(c) $U_a = 0.42 V_t$ and (b),(d) $U_a = 0.47 V_t$.
 }
 \label{fig:msd}
\end{figure}

The MSD was calculated for the particles in the bulk region.
Figure \ref{fig:msd} plots the MSD versus time $\tau$.
Assuming for the MSD the power-law form $\tau^{\beta}$, we see the ballistic regime $\beta \approx 1$ within a short time range, $\tau \lesssim \SI{0.1}{s}$.
During $0.1 < \tau < 0.3$ s the time dependence of the MSD changed rapidly to $\beta = 1/2$.
While in the case of $U_a = 0.47 V_t$ and $\phi_b \le 0.14$, the transitions were less clear.

In Fig.~\ref{fig:msd}cd, for clarity, the MSDs divided by $\tau$ are plotted so that the normal diffusion appear as zero gradient.
In the case of $U_a = 0.42 V_t$, as $\phi_b$ increased, $\beta$ gradually increased from less than 1/2 (subdiffusive) to slightly greater than 1/2 (superdiffusive).

Figure \ref{fig:msd} shows that the superdiffusive particles displaced about \SI{55}{mm} at the end of the time window $\tau = \SI{10}{s}$.
This displacement is about one-third of the diameter of the bulk region, and of 35 particles ($\phi_b=0.44$) that existed initially in the bulk region, on average 20 remained in the region after \SI{10}{s}.
To obtain statistically meaningful MSD, we have averaged over about 420 time windows (the data of \SI{2100}{s} total was splitted into 10-second windows with 50\% overlap).

\subsection{Velocity correlation function}

\begin{figure}[hbt!]
  \centering
  \includegraphics[width=\linewidth]{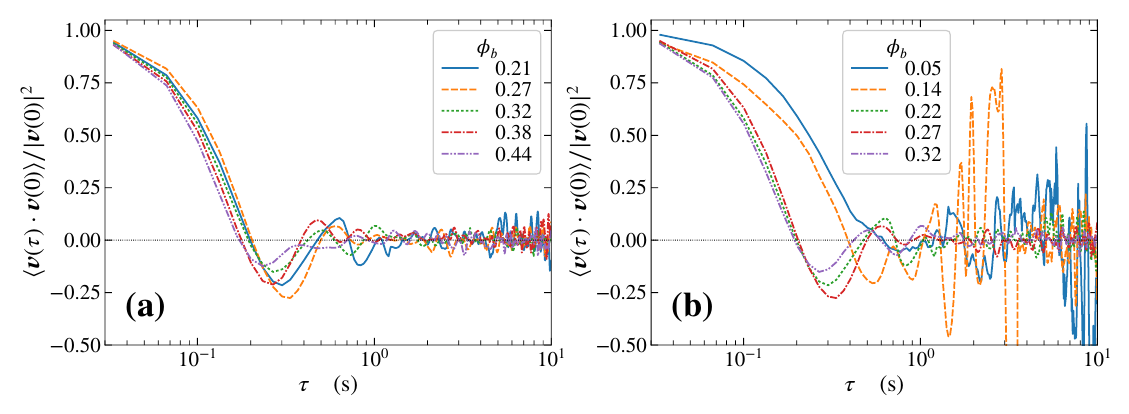}
  \caption{Velocity correlation functions for various $\phi_b$ at (a) $U_a = 0.42 V_t$ and (b) $U_a = 0.47 V_t$.
  The whole particle velocity data were divided into 10 s segments and the velocity autocorrelation for each segment was calculated and then averaged over 420 segments (with 50\% overlap) and over particles that did not exit the bulk throughout each time segment.}
  \label{fig:vcf}
\end{figure}

The velocity correlation functions for various $\phi_b$ are shown in Fig.~\ref{fig:vcf}.
Almost independently of $U_a$ and $\phi_b$, the correlation was almost lost within a short time $\tau \lesssim 1$ s with several oscillations, decaying into a background.
In the case of $U_a = 0.42 V_t$, the velocity correlation vanished at approximately \SI{0.2}{s} for all $\phi_b$.
The larger the $\phi_b$ was, the shorter the time interval between successive zeros and also the smaller the amplitude of the negative velocity correlation.

For $U_a = 0.47 V_t$, and small $\phi_b \lesssim 0.14$, the time of first zero correlation $\tau_0$ was longer than \SI{0.2}{s}.
On the other hand, when $\phi_b > 0.22$, $\tau_0 \approx \SI{0.2}{s}$ irrespective of $\phi_b$.
The intervals between successive zeros became shorter when $\phi_b$ was larger.
The oscillation of the correlation reached the largest when $\phi_b=0.27$.

\subsection{Velocity distribution function} \label{res:vdf}

\begin{figure}[hbt!]
  \centering
  \includegraphics[width=\linewidth]{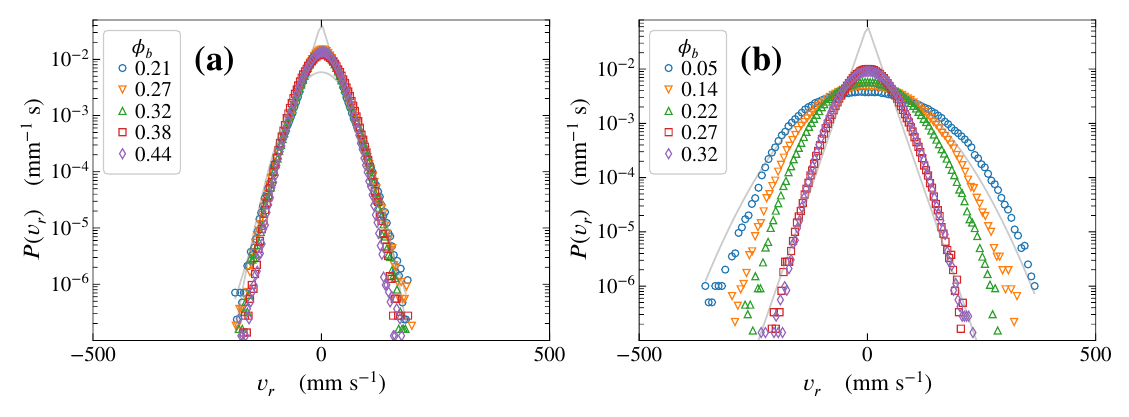}
  \caption{Radial velocity distribution functions $P(v_r)$ for various ball numbers $\phi_b$, at (a) $U_a = 0.42 V_t$ and (b) $U_a = 0.47 V_t$.
  Fitting the generalized exponential distribution $\exp(|v_r|^{-\alpha})$ to the data points gives $\alpha$ as a function of $U_a$ and $\phi_b$ as shown in Table \ref{tab:packingfraction}.
  Curves are given only as guides for the eye: $\alpha = 1.0$ (Cauchy; tent-shaped) and $2.0$ (Gaussian; parabola).
  }
  \label{fig:vdist_fit}
\end{figure}

The radial velocity distribution functions (VDF) for various $\phi_b$ are shown in Fig.~\ref{fig:vdist_fit} where the function
$C \exp(-D|v|^{\alpha})$
is fitted to the observed VDF (with $C,D,\alpha$ being the fitting parameters).
The values of $\alpha$ are listed in Table \ref{tab:packingfraction}.

For $U_a = 0.47 V_t$, in the case of fewer particles ($\phi_b \lesssim 0.14$), where no significant alignment was observed in the bulk region, the velocity distribution of the particles was close to the Gaussian distribution ($\alpha = 2$).
In contrast, for larger number of particles, the fraction of the particles that are significantly fast or slow was higher than that in Gaussian, so that the tail of the distribution became heavier. 
On the other hand, in the case of $U_a = 0.42 V_t$, $\alpha \approx 1.5$ for all $\phi_b$.
The kurtosis of the distribution slightly decreased in $0.38 < \phi_b <0.44$.

\section{Discussion}

\subsection{Effect of solid wall}

Previous laboratory studies on the air-fluidized granular materials have, to the best of our knowledge, implicitly limited their observation area to the bulk region, i.e. the central region detached from solid wall, focussing on homogeneous properties.
However, it is not obvious to what extent and how the presence of solid walls affects the motion of the particles in the bulk region.
In this subsection, we will examine the effect of walls. 

Figure \ref{fig:position} shows that a large number of particles gather selectively along the wall; the wall acts as an adsorption surface.
Although the aerodynamic mechanism of the wall adsorption is yet to be elucidated, it may be attributed to the frictional deceleration of the airflow (however, the viscous boundary layer at the top of the sieve wall is thinner than 3 \% of the particle diameter) and the modification of the wakes of the particles.
The particles in the adsorption layer acted collectively as a new adsorption surface, which lead to additional layers one after another, penetrating into the bulk region.
Such layering near the wall is frequently seen in the molecular dynamics simulations with solid boundaries \cite{md,Wang2018}.
The penetration length of the layer structure into the bulk region was longer for slower airflow, i.e. for weaker thermal agitation.
It is interesting to note that for $U_a = 0.42 V_t$, the layering reached a maximum at about $\phi_b=0.27$ and further increase of $\phi_b$ dissolved the layers.
The maximum penetration coincides with the occurrence of caging, which we will discuss in later section.

The presence of the wall also affected the thermal motion of the particles, in a localized way.
The ratio of the velocity variance in the direction perpendicular $\langle v^2_r \rangle$ to that parallel to the wall $\langle v^2_\theta \rangle$ is shown in Fig.~\ref{fig:thermalratio}.
Their ratios, which are zero at the wall, take maximum values greater than one at 1/2 particle-diameter from the wall and rapidly converge to unity within about 2 particle-diameters from the wall, regardless of the layer formation.
The degree of anisotropy of the thermal motion provides a handy quantity to make a distinction between the wall region (boundary layer) and the bulk region.

\subsection{Ballistic to diffusive transition}
\label{ss:transition}

In the long time scale ($\tau \gtrsim \SI{0.3}{s}$), the particles followed the random walk in which their MSD is approximately proportional to $\tau$ (Fig.~\ref{fig:msd}).
This can be regarded as the classical Brownian motion.
On the other hand, in the short time scale ($\tau \lesssim \SI{0.1}{s}$), the square root of MSD was proportional to $\tau$.

As seen in the velocity correlation function, the direction of the particle motion did not vary dramatically in the short time scale;  the particles are assumed to be in ballistic motion without particle-particle interaction.
The transition from ballistic to diffusive has been investigated using the Langevin equation with memory, in which the characteristic time for the viscous relaxation corresponds to the transition time.

The fluidized particles are driven by the random forces, which is the sum of the turbulent forces and the forces exerted by other particles due to inter-particle interactions.
From the Strouhal number obtained in a wind tunnel measurement \cite{Sakamoto1990}, the period of the wake of a sphere at Re $\approx$ 2700 can be estimated as $1/f \approx 0.05$ s.
This is sufficiently shorter than the time at which the velocity correlation function changes its sign, and thus it is considered that the inter-particle interactions were dominant in random forces that drive particles.

\subsection{Caging effect}

\begin{figure}[hbt!]
  \centering
  \includegraphics[width=\linewidth]{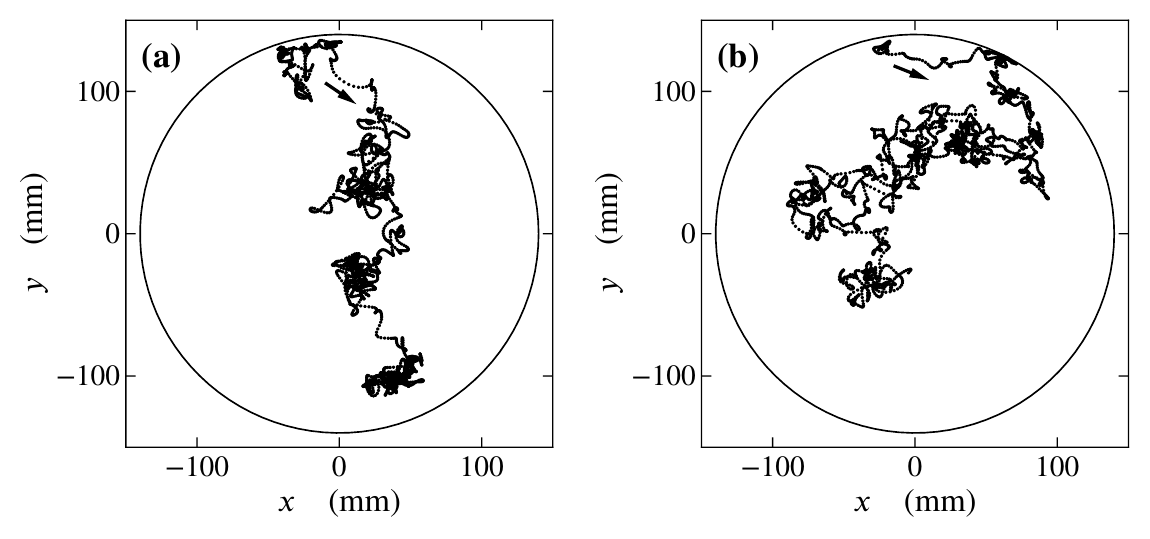}
 \caption{Sample trajectories of a particle near the center of the sieve at $U_a = 0.42 V_t$.
 Sample trajectories of a particle in a interval of \SI{100}{s} and for (a) $\phi_b \approx 0.32$ 
 and (b) $\phi_b \approx 0.44$ are shown with time interval of \SI{1/30}{s} between the dots.
 The arrows show the direction of particle movement.
 The circles designate the effective sieve boundary at radius $R_{\textrm{sieve}}-R_{\textrm{particle}} = \SI{140}{mm}$.}
 \label{fig:tracking}
\end{figure}

In the case of the particle alignment in the bulk region ($0.27 < \phi_b< 0.32$), a plateau appeared on the MSD immediately after the ballistic-diffusive transition.
Such behavior is often observed in glass and dense colloidal systems.
This phenomenon is considered to result from stable enclosing by surrounding particles and is known as the caging effect \cite{Zaccarelli2009} and has been observed also in granular experiments \cite{Abate2006,Keys2007,Abate2007}.
It is noteworthy that in the same range of $\phi_b$ the particle alignment was seen (Fig. \ref{fig:position}).    
Furthermore, in this range, particle trajectory appeared as classical random walk interspersed with arrests, indicating the caging effect \cite{Reis2007}.
A typical trajectory at $U_a = 0.42 V_t ,\; \phi_b = 0.32$ is shown in Fig. \ref{fig:tracking}a.
Such trajectories and the MSD plateau, both of which are indicative of enhanced confinement due to caging, were more pronounced at slower airflow.

The most remarkable result is that the present particle system showed the characteristics of caging at a significantly lower packing fraction than those in previous experiments.
Hard spheres exhibit caging at $\phi \approx 0.84$ \cite{Philipse2003}, which is very close to the random packing limit.
For a vibrated monolayer of steel balls \cite{Reis2007}, a plateau appears in the MSD at $\phi \approx 0.7$.
It is reported that the caging occurred at $\phi \approx 0.7$ for fluidized ping-pong balls ($\rho = \SI{0.08}{g.cm^{-3}}$) \cite{Lopezcastano2019}, while for fluidized beads no caging was observed \cite{Beverland2011}.
Fluidized bidispersed steel balls \cite{Abate2006} exhibit strong caging at $\phi = 0.74$ but do not show subdiffusion for $\phi < 0.64$.
In contrast, we observed in this experiment a caging plateau at $\phi_b \approx 0.32$.
This is arguably due to the low material density of the particles, 
between which the predominant interaction was switched from direct collision to indirect interaction via the airflow.
The switching is also seen in the particle-radial distribution functions (RDFs) \citesupplementalmaterial, where only the most sparse cases show the gas-like RDF.
It is also noticed that at the slower air speed the range of repulsion became wider hence the repulsive potential was softer,
which may be related to that the caging occured at lower $\phi_b$ than heavier particles.

\begin{figure}[hbt!]
  \centering
  \includegraphics[width=\linewidth]{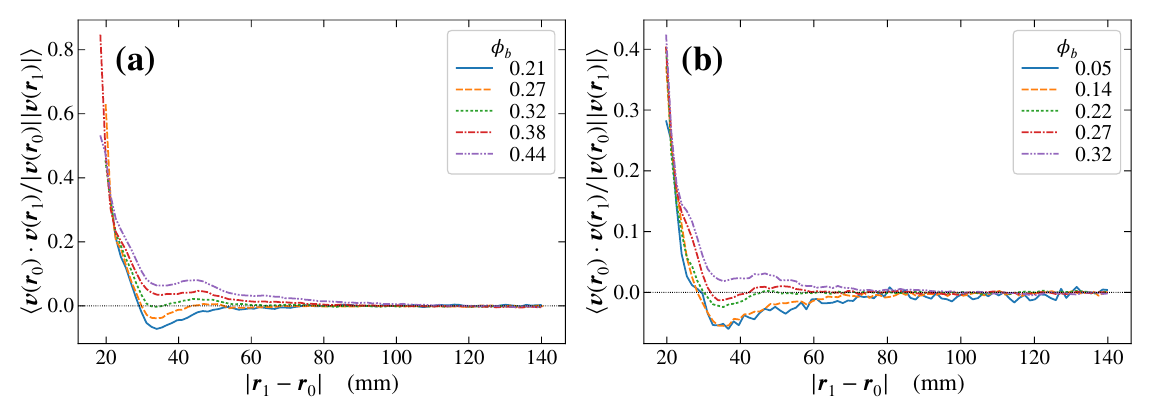}
  \caption{Two-body velocity correlation functions $\langle \bm{v}_0\cdot \bm{v}_1 \rangle/|\bm{v}_0||\bm{v}_1|$,
     for various $\phi_b$, at (a) $U_a = 0.42 V_t$ and (b) $U_a = 0.47 V_t$.
     The segmentation of the time series and the averaging were performed in the same way as Fig.~\ref{fig:vcf}.}
  \label{fig:tcf}
\end{figure}

\subsection{Anomalous diffusion}

At $U_a = 0.42 V_t$, when $\phi_b$ was further increased after the particle alignment was observed, the plateau after the transition disappeared and the particle alignment was no longer observed.
This is considered to correspond to that the particle arrangement became obscure (Fig.~\ref{fig:position}).
Furthermore, at the same time, the time exponent of the particle displacement became $\beta > 1/2$ (superdiffusive).
The general mechanism of such anomalous diffusion is yet to be fully understood.
As an attempt to find a factor that caused the anomalous diffusion, we investigated the inter-particle correlation for the particle motion.
We show in Fig.~\ref{fig:tcf} the two-body velocity correlation, which provides a measure of the kinematic similarity between the focused particle and neighboring particles.

At any airspeed, correlation length, which is the characteristic length of the non-zero two-body velocity correlation, became longer with the increase of $\phi_b$.
The correlation length dramatically extended at around $\phi_b \approx 0.32$.
When $\phi_b$ was large, particle motion became spatially coherent, and anomalous diffusion of the particle system occurred.

The fact that the size of the system is only $N \sim O(10^2)$ in this experiment requires further discussion.
In the anomalous diffusion discussed in this subsection, it cannot be ruled out that there is some finite size effect.
Thus the findings here might not be generalizable to infinitely large systems.
Rather, it seems more appropriate to point out that the present experiment has more relevance to colloids in porous media \cite{Molnar2015} and in nanofluidic devices \cite{Wang2018}.

\subsection{Velocity distribution function}
A caged particle undergoes more confined motion than the classical thermal motion.
This leads to more particles with small speed ($v \approx 0$), and to the velocity distribution function with a heavier tail than the Gaussian distribution.
The fitting function $\Phi=C\exp(-D|x|^{\alpha})$ for velocity distribution function can also be regarded as the characteristic function for probability distribution of displacement, where Gaussian distribution and Cauchy distribution correspond to $\alpha=2$ and $\alpha=1$ respectively \cite{Gardiner2004}.

In the case of $U_a = 0.42 V_t$, $\alpha$ became larger as the structure relaxed when $\phi_b \lesssim 0.38$, while $\alpha$ decreased when $\phi_b \approx 0.44$.
It is considered that this resulted from the crowding of particles that prevents the particle from free motion at large $\phi_b$.
In the case of $U_a = 0.42 V_t$, $\alpha$ took intermediate value between the Gaussian distribution and the Cauchy distribution for all $\phi_b$.
On the other hand, in the case of $U_a = 0.47 V_t$, the velocity distribution function became a heavier tail distribution when $\phi_b$ increased.
It approached to the Cauchy distribution, but remained $\alpha>1$.

\subsection{Aerodynamic forces}

Now we discuss the forces driving the particles, i.e. the aerodynamic forces.
The flow acting upon the particle (mainstream) is basically laminar because it has just passed through the fine mesh of the sieve and thus the flow is basically laminar \cite{Groth1988}. Therefore, it is the turbulence created by the particle itself that drives it randomly.

The flow behind the particle was measured using a hot wire anemometer \citesupplementalmaterial.
Calculating aerodynamic force acting on the particle requires stresses over the entire surface, which is technically difficult to measure.
Instead, we estimated the sieve-radial component $F_r$ of the aerodynamic force assuming the airflow immediately above the particle as representative of the fluctuating flow responsible for the forcing.
For the present flow Reynolds number is large, thus the aerodynamic force exerted by a fluid is mostly pressure, which is proportional to the square of the flow velocity.

\begin{figure}[hbt!]
  \centering
  \includegraphics[width=\linewidth]{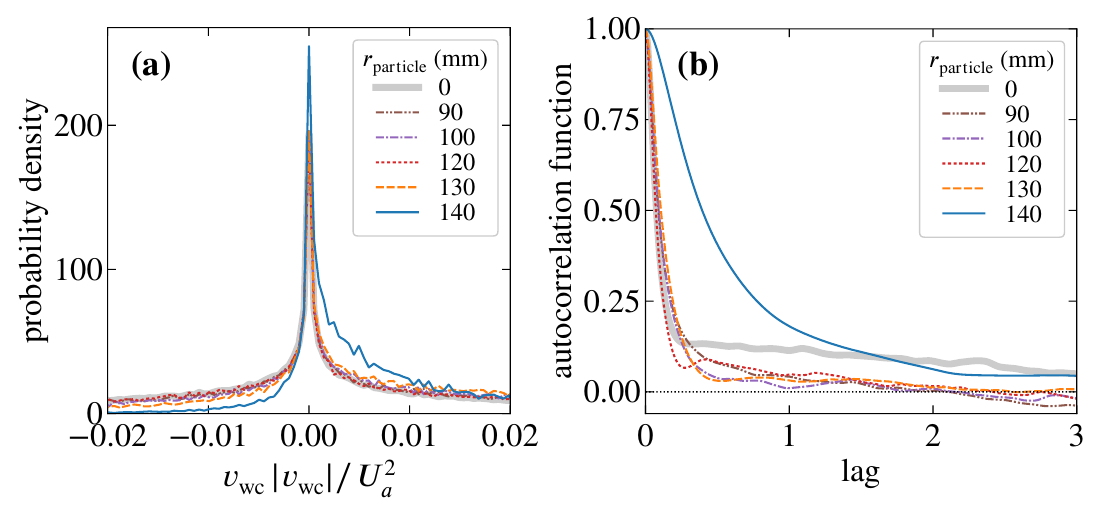}
  \caption{(a) Histogram of aerodynamic force in the wake of single spherical particle rigidly fixed to the sieve surface at various radial positions in sieve $r_{\mathrm{particle}}$.
  The sieve-radial velocity $v_{\mathrm{wc}}$, which was measured at 5 mm downstream along wake centerline, is squared to be proportional to its contribution to aerodynamic force, and is normalized by $\overline{u}^2$.
  Positive values on the horizontal axis are forces toward the wall and negative values are toward the center.
  (b) Autocorrelation of $v_{\mathrm{wc}}$.  The lag is normalized by the characteristic time scale for the flow past a particle, $R_{\mathrm{particle}}/\overline{u}$.}
  \label{fig:turbulence}
\end{figure}

Figure \ref{fig:turbulence}a shows the histogram of estimated $F_r$.
We can see that $F_r$ is symmetric and invariant in the bulk region, but as the particle approaches the wall, the distribution skews significantly to the right.
$F_r$ is, as shown in Fig.~\ref{fig:turbulence}b, highly fluctuating force, which is coherent only over a fraction of aerodynamical characteristic time, $R_{\mathrm{particle}}/\overline{u}$,
except when the particle's contact to the wall directly obstructs the flow.
This short timescale corresponds to that of the vortex shedding of the sphere \cite{Sakamoto1990},
which has been discussed in Section \ref{ss:transition}.
This timescale is about $10^2$ times smaller than the observed relaxation time of the particles (Fig.~\ref{fig:vcf}).
The rapid decorrelation of the forcing compared to the particle motion also suggests that the forcing is effectively Markovian, which has been well confirmed in turbulent wakes \cite{Tutkun2004,Siefert2006}.
Therefore we are lead to a picture that the aerodynamic ``kick'' is homogeneous throughout the bulk, while in the vicinity of the wall the ``kick'' becomes increasingly wall-ward.
In the thermal analogy, the aerodynamic forcing due to the multitude of wake vortices can be interpreted as a thermal bath surrounded by a cold wall that traps the particles.

By extending the above estimation with a dimensional argument, we can also deduce the effect of ``lightness'' of particles on their thermal motion.
The aerodynamic impulses exerted upon a particle, $\rho_{\mathrm{a}} (2R_{\mathrm{particle}})^2 U_a^2 (2R_{\mathrm{particle}}/U_a) = \rho_{\mathrm{a}} (2R_{\mathrm{particle}})^3 U_a$, where $\rho_{\mathrm{a}}$ is the air density, lead to the momentum change, $\rho_{\mathrm{p}} (2R_{\mathrm{particle}})^3 V_p$, where $\rho_{\mathrm{p}}$ is the particle's material density and $V_p$ the change of the particle velocity due to the impulses.
Thus $V_p / U_a \propto \rho_{\mathrm{a}}/\rho_{\mathrm{p}}$.
This means that, as expected, the lighter the particle, the more susceptible the particle's motion is to turbulence.

Finally, consider the limit of vanishing material density of particle.
The airflow would have to be extremely slow to keep the particles from being blown away.
In such a low Reynolds number flow, spherical particle pair potential can be attractive \cite{Folkersma2000}, leading to the aggregation.
This suggests phenomena such as shear localization seen in foam and bubble rafts \cite{Debregeas2001,Lauridsen2004}, which are typical systems of aggregates with small inertia, may be observed in systems of more lightweight fluidized particles.




\section{Conclusion}

We have presented the statistical characteristics of the motion of two-dimensional gas-fluidized lightweight particles derived by image analysis techniques.
The main findings are as follows.

The behavior of the particles was largely different between in the wall region and in the bulk region, where the particle motions were anisotropic and isotropic, respectively.
When $\phi_b$ was increased with the airspeed fixed, the ordered structure of the particle was formed from the wall and gradually penetrated into the bulk region.
If compared at the same $\phi_b$, the particles were less ordered in faster airflow.
As the particle ordering penetrated into the bulk region, a plateau began to appear on the plot of the MSD as in the similar systems previously reported. 

However, when $\phi_b$ was increased beyond the point of bulk ordering, the structure became increasingly disordered and the MSD exhibited superdiffusion.
At this stage, the two-body velocity correlation showed that the correlation length became longer for larger $\phi_b$.
These features may suggest the formation of the heterogeneous structures which is a mosaic of jammed clusters and fast-moving stringy clusters.
As a further step in this direction, experiments in much larger system will be considered next.
They are reserved for future work.

Another remarkable aspect of the lightness of the particle is that all these transitions occurred at much lower $\phi_b$ than those reported previously.  The radial distribution functions also exhibited the gas-like to liquid-like transition.
They may imply that by using lightweight particles we can emulate a soft potential in the context of fluidized particle experiments.
The air-fluidization provides a simple means for emulating thermal agitation when simulating molecular kinetics using macroscopic particles.

\begin{acknowledgments}
Part of the experimental apparatus were kindly provided by Satoshi Sakai and Naoto Yokoyama.
We thank Shigeru Takata for useful comments, the participants of The Tottori Nonlinearity Workshop for instructive discussions.
Thanks are extended to Shogo Hashi and Michihiro Miyashita for their assistance in the hot-wire anemometry.

\end{acknowledgments}

%

\end{document}